\documentclass[twocolumn,prb,amsmath,amssymb,unsortedaddress,showpacs,floatfix,citeautoscript,nofootinbib]{revtex4-1}

\usepackage{latexsym,epsfig,graphicx}
\usepackage{dcolumn}
\usepackage{bm}
\usepackage{graphicx}
\usepackage{subfigure}
\usepackage{color}
\usepackage[colorlinks,urlcolor=blue,citecolor=blue]{hyperref}

\begin{document}

\title{Bose-Hubbard model with occupation-parity couplings}
\author{Kuei Sun}
\author{C. J. Bolech}\affiliation {Department of Physics, University of Cincinnati, Cincinnati,
Ohio 45221-0011, USA}

\date{February 18, 2014}


\begin{abstract}
We study a Bose-Hubbard model having on-site repulsion,
nearest-neighbor tunneling, and ferromagneticlike coupling between
occupation parities of nearest-neighbor sites. For a uniform
system in any dimension at zero tunneling, we obtain an exact
phase diagram characterized by Mott-insulator (MI) and pair liquid
phases and regions of phase separation of two MIs. For a general
trapped system in one and two dimensions with finite tunneling, we
perform quantum Monte Carlo and Gutzwiller mean-field
calculations, both of which show the evolution of the system, as
the parity coupling increases, from a superfluid to
wedding-cake-structure MIs with their occupations jumping by 2. We
also identify an exotic pair superfluid at relatively large
tunneling strength. Our model ought to effectively describe recent
findings in imbalanced Fermi gases in two-dimensional optical
lattices and also potentially apply to an anisotropic version of
bilinear-biquadratic spin systems.
\end{abstract}

\maketitle

\section{Introduction}\label{sec:intro}

Since half a century ago, the Hubbard model has been widely
studied as a simple model accounting for several nontrivial
phenomena in electronic systems, such as the Mott insulator,
(anti)ferromagnetism, and novel
superconductivity\cite{Hubbard63,Belitz94,Dagotto94,Imada98,Lee06}.
Its bosonic version, the Bose-Hubbard (BH) model, has been
appreciably investigated for more than two
decades\cite{Fisher89,Sachdev99}. The BH model not only aptly
describes the many-body behavior of bosons in lattices, but also
serves as a good and simple example for understanding how
competition between two common mechanisms, localization and
itineracy, can drive matter toward extremely different phases,
namely Mott-insulator (MI) or superfluid. The MI phase,
characterized by commensurate occupations, gapped excitations and
incompressibility, undergoes a transition to a superfluid,
characterized by Bose-Einstein condensation, gapless excitations
and finite compressibility, as the ratio of tunneling to
interaction energy increases beyond a density-dependent critical
value. Differently from the fermionic case, the bosonic nature
allows the BH phase diagram of multiple MI regions for all
possible integer occupations, surrounded by the superfluid region.
Various realizations of the BH model have been early suggested in
granular systems with an embedded condensate, such as a Josephson
junction array\cite{Zant92,Oudenaarden96} or liquid helium in
porous media\cite{Dimeo98,Plantevin01}, and later in lattice
bosons, such as bosonic atoms in optical lattices\cite{Jaksch98}
or exciton-polaritons in an array of
microcavities\cite{Hartmann06,Greentree06,Lai07,Byrnes10}. Recent
achievements in cold-atom experiments have successfully created
inhomogeneous BH systems and realized the MI-superfluid
transition\cite{Greiner02} as well as a spatially separated
structure of multiple MIs\cite{Campbell06}. These multiple
applications have stimulated a broad interest in the BH physics
and lead to an active study of a great variety of BH models
\cite{Jaksch05,Morsch06,Bloch08,Cazalilla11,Marzari12}.

Recently, a theoretical investigation we carried out on imbalanced
fermionic superfluids in a two-dimensional (2D) array of coupled
tubes\cite{Sun12} [a setup which has been experimentally
realized\cite{Liao10} in the search for the elusive
Fulde-Ferrell-Larkin-Ovchinnikov (FFLO)
state\cite{FF,LO,Giorgini08,Radzihovsky10}] has shown an exotic
compressible-incompressible quantum phase transition of the tube
occupation of unpaired majority fermions (UMFs) in the system. The
phase diagram obtained from the microscopic model of fermions
exhibits similarities and contrasts to the structure as a BH phase
diagram. On the one hand there is a phase with well defined UMF
occupation numbers in each tube. On the other hand, the
occupations of the incompressible (MI) regions are either all even
or all odd. The similarities reflect the competition between the
ontube energy and cross-tube kinetics of the UMFs and hence
suggests an effective description of the (projected) system by a
2D BH model, while the contrasts suggest the need for an
additional term in the effective model that ``orients'' the
occupation parities (see Sec.~\ref{sec:model} below for a detailed
discussion).

\begin{figure}[t]
\centering
   \includegraphics[width=8cm]{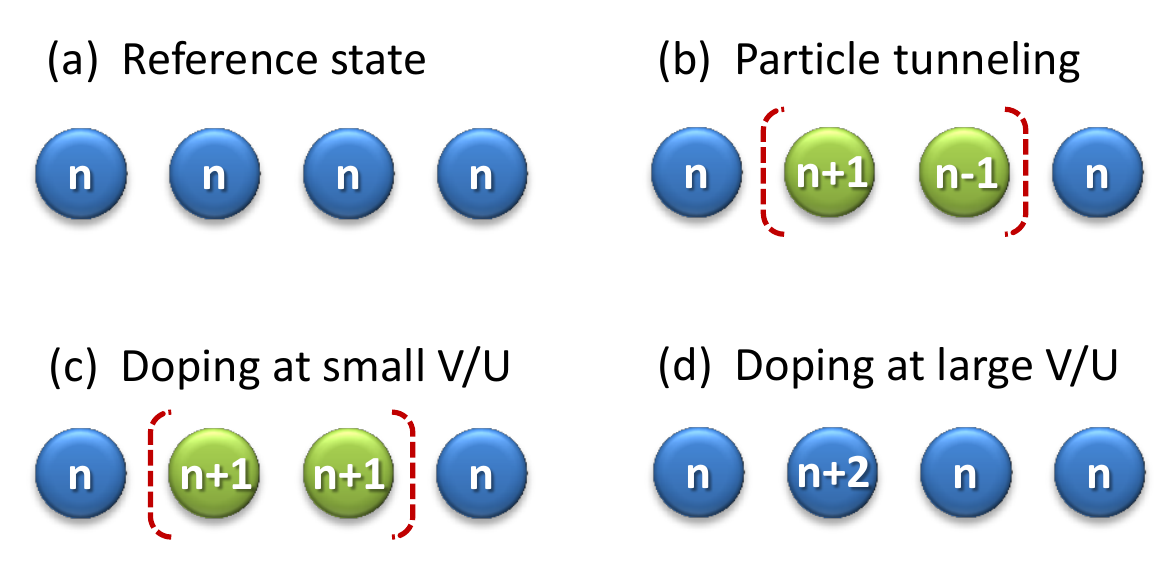}
  \caption{(Color online) Illustration of the energetic competition between
  particle tunneling $t$, interaction $U$, and parity coupling $V$ of the
  system. (a)
  A reference state of each site occupied by $n$ particles. (b) After a
  single particle tunneling, the change in number of two neighbor sites
  alters the interaction energy by $O(U)$,
  while the change in parity of them results
  in a domain-wall formation (dashed curve) of energy cost $O(V)$.
  (c) At small $V/U$, two doping particles tend to occupy different sites,
  minimizing the interaction energy. (d) At large
  $V/U$, the two occupy the same site as a pair doping,
  avoiding the domain-wall energy.}
        \label{fig:f01}
        \vspace{-0.5cm}
\end{figure}

Motivated by this previous work, we study an extended BH model
having a ferromagneticlike coupling between occupation parities of
nearest-neighbor sites. This proposed coupling can be pictured as
domain-wall energies between nearest-neighbor sites of opposite
occupation parities, reminiscent of local magnetization kinks in
the ferromagnetic Ising model\cite{Sachdev99}, resulting in a
ground state that favors all the sites having the same occupation
parity. The parity coupling shows interesting interplays with the
other two original BH ingredients, single-particle tunneling and
onsite repulsion. First, it is antagonistic toward the
single-particle tunneling process, during which two involved
neighbor sites flip the parity and form domain walls between
themselves and the surrounding sites [see illustration in
Figs.~\ref{fig:f01}(a) and \ref{fig:f01}(b)]. Second, it is
different from the onsite repulsion in that it drives the system
to accept additional doping as pairs rather than single particles,
to avoid the domain wall energy, while the onsite repulsion does
the opposite as a way to reduce the interaction energy [see
Fig.~\ref{fig:f01}(c) and \ref{fig:f01}(d)]. In this paper, we
analyze the system and characterize the effects of the
occupation-parity coupling. We will suggest two possible
experimental realizations for exploring the model in cold atomic
and condensed matter systems. We shall find a rich phase diagram
for the uniform case in the zero-tunneling limit, reflecting
different doping preferences driven by the parity coupling and
onsite repulsion. For the finite-tunneling case, we apply two
commonly used treatments, quantum Monte
Carlo\cite{Alet05b,Alet05a,Albuquerque07,Bauer11} (QMC) and
Gutzwiller mean-field\cite{Rokhsar91} (GMF) methods, on trapped
systems, thus successfully describing the state of the system as
the parity coupling increases for finite tunneling. In the
large-tunneling regime, we identify an exotic pair superfluid
state, which emerges with a different mechanism from previously
studied
ones\cite{Trefzger09,Hu09,Menotti10,Diehl10,Bonnes11,Ng11,Chen11}.

The paper outline is as follows. In Sec.~\ref{sec:model}, we
introduce the model Hamiltonian and discuss its possible
realizations by studying its emergence in imbalance Fermi gases as
well as in spin systems. In Sec.~\ref{sec:phase}, we perform a
detailed analysis of the competition between the parity coupling
and the onsite repulsion, and obtain an exact phase diagram in the
zero-tunneling plane. In Sec.~\ref{sec:numerical}, we present
results from QMC and GMF calculations that show the evolution of a
trapped system and the emergence of the pair superfluid. Finally,
we summarize our work in Sec.~\ref{sec:conclusion}.

\section{Model}\label{sec:model}

The extended BH model with occupation-parity couplings between
nearest-neighbor sites is described by the Hamiltonian
\begin{eqnarray}
H &=& \sum\limits_{\left\langle {ij} \right\rangle } { - t(\hat
b_i^\dag {{\hat b}_j} + {\rm{H}}{\rm{.c}}{\rm{.}}) -
\frac{V}{2}({{\hat P}_i}{{\hat P}_j} - 1)} \nonumber\\
&& + \sum\limits_i {\frac{U}{2}{{\hat n}_i}({{\hat n}_i} - 1) -
{\mu _i}{{\hat n}_i}}  \label{eqn:Ham0},
\end{eqnarray}
where $\hat b_i$ is a bosonic operator on site $i$, $\hat n_i =
\hat b_i^\dag \hat b_i$ is the number operator, $\hat P_i =
(-1)^{\hat n_i}$ is the number-parity operator, $\mu_i$ is the
local chemical potential and $\left\langle {ij} \right\rangle$
denotes a pair of neighboring sites. The non-negative parameters
$t$, $V$, and $U$ give the strength of nearest-neighbor tunneling,
ferromagnetic-like nearest-neighbor parity coupling and onsite
repulsion, respectively. Each pair of them show a competition that
is reflected in the ground state of the system. At $V=0$, the
Hamiltonian returns to the original BH one, in which the
domination of itineracy (large $t/U$) or localization (large
$U/t$) results in a superfluid or MI, respectively. The presence
of $V$ terms can be regarded as an energy cost of domain walls
between two neighboring sites with different number parities. From
this point of view, the competition between $t$ and $V$ can be
described by the picture that a single-particle tunneling changes
the parities of two sites and hence pays an energy cost of
creating a domain wall surrounding them [see
Figs.~\ref{fig:f01}(a) and \ref{fig:f01}(b)]. Therefore, when $V$
dominates, the system favors the minimization of the total number
of domain walls, that is, to have all sites with the same
occupation parity. The interplay between $U$ and $V$ yields a rich
phase diagram, which will be discussed in detail in
Sec.~\ref{sec:phase}. Here, we now turn to discuss two possible
realizations of our model.

First, a recent study in Ref.~[\onlinecite{Sun12}] has suggested
that the mechanism in the Hamiltonian of Eq.~(\ref{eqn:Ham0})
accounts for an exotic quantum phase transition in imbalanced
fermionic superfluids in optical lattices. The system has the
geometry of a 2D array composed of one-dimensional (1D) tubes that
has been realized in
experiments\cite{Moritz05,Liao10,Aidelsburger11}. In such a system
it is the behavior of spin imbalance per tube, or the fillings of
UMFs, that undergoes a transition from a compressible state (UMF
move easily across tubes) to a MI (UMFs are localized on each
tube), yielding a similar phase diagram to the regular BH one
except the filling numbers of the MI phases are either all even or
all odd while they can be any integer in the regular case.
Regarding the projection of this anisotropic three-dimensional
fermionic system to an effective 2D lattice model described by
Eq.~(\ref{eqn:Ham0}), we find the tunneling $t$ and on-site
interaction $U$ as a result from the interplay of the lattice
geometry, on-tube interaction, and on-tube pairing order. The
occupation-parity coupling $V$ comes from the domain-wall energy
between two neighboring tubes having different spatial parities of
the oscillatory superconducting order parameter, which is directly
related to the occupation parity of UMFs. In the Appendix
\ref{sec:DBH}, we present a detailed derivation showing how the
physics of UMFs is effectively described by Eq.~(\ref{eqn:Ham0}),
by starting from the original Hamiltonian of the fermionic
superfluid system. We remark that the parameter regimes discussed
in Ref.~[\onlinecite{Sun12}] are for the large $V$ limit. However,
the high tunability of lattice geometry and interatomic
interaction in this cold atomic system could enable the
exploration of sufficiently wide parameter regimes of our model
Hamiltonian.

Second, the regular BH model can be mapped to a spin-1
(${{{\bf{\hat S}}}}$) system around the tip of the Mott insulator
lobe in the large-occupation limit, in which a three-state
truncation applies on each site\cite{Altman02}. For our model, the
occupation parity coupling terms are mapped to spin couplings of
the Ising-type with quadratic and biquadratic forms. Taking the
truncation basis of integer occupation states $\left| n
\right\rangle$ and $\left| n \pm 1 \right\rangle$, we perform the
mapping ${{\hat b}^\dag } \to \sqrt n {{\hat S}^ + }$, $\hat n \to
n + {{\hat S}^z}$ as well as $\hat P \to {( - 1)^n}[ {1 - 2{{(
{{{\hat S}^z}})}^2}} ]$ on Eq.~(\ref{eqn:Ham0}) and obtain a spin
Hamiltonian,
\begin{eqnarray}
{H_{{\rm{S}}}} &=& \sum\limits_{\left\langle {ij} \right\rangle }
{ - {J_1}\left( {\hat S_i^x\hat S_j^x + \hat S_i^y\hat S_j^y}
\right) - {J_2}{{\left( {\hat S_i^z} \right)}^2}{{\left( {\hat
S_j^z} \right)}^2}}  \nonumber\\
&&+ \sum\limits_i {D{{\left( {\hat S_i^z} \right)}^2} + {h_i}\hat
S_i^z}, \label{eqn:Ham1}
\end{eqnarray}
up to a constant offset. Here $J_1=2nt$, $J_2=2V$, $D=ZV+U/2$ with
$Z$ being the coordination number and $h_i=(n-1/2)U-\mu_i$. One
immediately sees the competitions in Eq.~(\ref{eqn:Ham1}): the
bilinear coupling $J_1$ aligns spins in the $x$-$y$ plane, the
biquadratic coupling $J_2$ drives spins to $S^z=\pm1$ states, the
non-linear Zeeman coupling $D$ drives them to $S^z=0$ state, and
the local magnetic field $h_i$ aligns them in the $z$ direction.
The first line in Eq.~(\ref{eqn:Ham1}) can be regarded as an
anisotropic version of standard bilinear-biquadratic spin
models\cite{Barber89,Fath91,Ivanov03,Legeza06,Lauchli06,Liu12,Fridman13},
whose experimental realization has been discussed in condensed
matter\cite{Tsunetsugu06} and cold atomic
gases\cite{Yip03,Garcia-Ripoll04}. The second line commonly
appears in nonlinear spin
systems\cite{Haldane83,Kadowaki87,Affleck89}. A recent study on
fermionic dipolar gases\cite{Kestner11} also found a route toward
the creation of an anisotropic spin system, as a variant of
Eq.~(\ref{eqn:Ham1}). Due to the connection between the two
models, the realization of the spin system could provide an
alternative approach for examining a variant of our model bosonic
system, and vice versa.

\section{Phase diagram at $t=0$}\label{sec:phase}

In this section we discuss the competition between the onsite
interaction $U$ and the parity coupling $V$ through additional
doping of a uniform system in the zero-tunneling regime. We derive
exact energy functionals in any dimension for each competing phase
and obtain the ground-state phase diagram marking their relatively
energetically favorable regions.

At $t=0$, the ground state is a product of each single-site state
with integer occupation $n_i$ on site $i$. The configuration of
$\{n_i\}$ is obtained by minimizing the system's energy. At
integer fillings $n$, we have a MI with all $n_i=n$
[Fig.~\ref{fig:f01}(a)], independent of the ratio $V/U$. Now
considering the doping of two additional particles in the system,
if $U$ dominates (small $V/U$), we expect that the two particles
will occupy different sites to avoid higher interaction energy
[Fig.~\ref{fig:f01}(c)]. Such two sites with $n+1$ occupation can
be arbitrary ones at $V=0$ and should bind together as neighbors
at $V>0$ due to the \emph{domain-wall} effect. For more than two
particles added, they similarly tend to singly dope sites that
randomly spread if $V=0$ or cluster if $V>0$. We call the former
case a single-particle doped liquid (SL) and the latter a phase
separation of two MIs (PS). If $V$ dominates (large $V/U$), to
avoid the domain wall energies, the two doping particles tend to
occupy the same site producing pair doping
[Fig.~\ref{fig:f01}(d)], thus keeping the parity of all sites
unchanged. More than two particles will doubly dope in the same
way, up to one doping pair per site due to the interaction
effects. We call such state a pair-doped liquid (PL). Similarly,
when particles are taken out of the system (in analogy to the hole
doping in fermionic systems), they leave as single particles at
large $U$ and as onsite pairs at large $V$. Below we calculate the
phase boundary between PS and PL in a homogeneous system.

In the case of doping $M_\pm$ particles (holes) in a
$d$-dimensional uniform system of $L^d$ sites with integer filling
$n$ ($M_\pm$ is taken even and $M \le L^d/2$), the PS state has
$M_\pm$ sites of $n \pm 1$ filling, $L-M_\pm$ sites of $n$ filling
and $N_{\rm{DW}}$ domain-wall links, while the PL state has
$M_\pm/2$ sites of $n\pm 2$ filling, $L-M_\pm /2$ sites of $n$
filling, and no domain walls. The energies of the PS and PL states
can thus be written as
\begin{eqnarray}
{E_{{\rm{PS}}}}(n,{M_ \pm }) &=& {M_ \pm }{E_U}(n \pm 1)\nonumber\\
&& + ({L^d} - {M_ \pm }){E_U}(n) + {N_{\rm{DW}}}V,\label{eqn:EPS}
\end{eqnarray}
and
\begin{eqnarray}
{E_{{\rm{PL}}}}(n,{M_ \pm }) &=& \frac{{{M_ \pm }}}{2}{E_U}(n \pm
2) \nonumber\\
&& + \left({L^d} - \frac{{{M_ \pm
}}}{2}\right){E_U}(n)\label{eqn:EPL},
\end{eqnarray}
respectively, where $E_U(n)=Un(n-1)/2$ is the onsite interaction
energy. We let ${E_{{\rm{PS}}}} = {E_{{\rm{PL}}}}$ to obtain a
critical ratio $(V/U)_c$ that the separates PS and PL states,
\begin{eqnarray}
{\left( {\frac{V}{U}} \right)_c} = \frac{{{M_ \pm
}}}{{2{N_{\rm{DW}}}}}.\label{eqn:VUc}
\end{eqnarray}
Given periodic boundary conditions for the system, the doped
domain favors a spherical shape to minimize the surface area or
the number of domain-wall links, so we have ${M_ \pm } = {\tau
_d}{R^d}$ and ${N_{\rm{DW}}} = {s_d}{R^{d - 1}}$, where $R$ is the
domain radius and $\tau_d$ ($s_d$) is the volume (surface area) of
a $d$-dimensional unit sphere. Substituting these relations in
Eq.~(\ref{eqn:VUc}) and rewriting the doping in terms of a
rescaled coupling $\tilde V= V/(UL)$ and the average filling $\bar
n$ such that ${M_ \pm } = \left| {\bar n - n} \right|{L^d}$, we
obtain
\begin{eqnarray}
\tilde V_c = {A_d}{\left| {\bar n - n}
\right|^{1/d}},\label{eqn:VUc2}
\end{eqnarray}
where ${A_d} = {({\tau _d})^{\left( {1 - 1/d} \right)}}/(2{s_d})$.
Equation (\ref{eqn:VUc2}) for $\bar n -n <1/2$ (respectively,
$-1/2< \bar n -n$) describes the PS-PL transition due to particle
(hole) doping on the $n$-filling MI; except for $0<\bar n <1$ when
there is no PL that is hole doped from the $n=1$ MI, so the phases
are determined only by particle doping on the vacuum state, where
$\tilde V_c = {A_d}{{\bar n}^{1/d}}$.

\begin{figure}[t]
\centering
   \includegraphics[width=7.5cm]{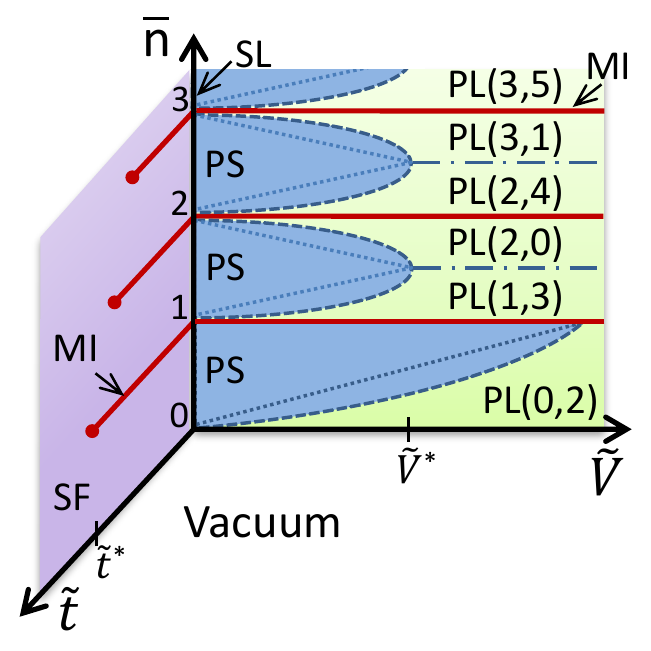}
  \caption{(Color online) Schematic phase diagram combining $V=0$ and $t=0$
  cases. Here $\tilde t=t/U$ and $\tilde V=V/UL$ with $L^d$ being the volume
  of the system (in $d$ dimensions). In the $V=0$ plane, the system is a Mott
  insulator (MI, solid section line)
  at integer fillings and $\tilde t \le \tilde t^*(d,n)$ or is a superfluid (SF)
  otherwise. In the $t=0$ plane, the system is MI (solid line)
  at integer fillings and any $V$. At any fractional filling and finite $V$, the system
  is phase separated into two MIs (PS), or is a pair-doped liquid (PL), distinguished by the critical relation in
  Eq.~(\ref{eqn:VUc2}) [dashed (dotted) curve in two (one) dimensions]. In the PL
  regime at $1 \le n <\bar n < n+1$, the half-integer fillings (dash-dotted
  curve) separate two types of PL of different parities,
  PL$(n,n+2)$ as well as PL$(n+1,n-1)$, and meet the PS lobe at a
  ``triple point" $\tilde V^*(d)$ (see text for more details). In the $\bar n$ axis, the fractional-filling
  region is a single-particle doped liquid (SL).}
        \label{fig:f02}
        \vspace{-0.5cm}
\end{figure}

Figure \ref{fig:f02} shows a combined phase diagram of the $V=0$
($\bar n$--$\tilde t$ plane, where $\tilde t=t/U$) and $t=0$
($\bar n$--$\tilde V$ plane) cases in one and two dimensions. At
$V=0$, the system turns to the original BH model: the ground state
is a MI (solid section lines) at integer fillings and $\tilde t$
below a critical value $\tilde t^*$ or is a superfluid (SF) at any
fractional filling. Here $\tilde t^*$ monotonically increases with
dimension $d$ and decreases with filling $n$. At $t=0$, the
integer fillings are always MIs (the solid lines extend to $V \to
\infty$), while at fractional fillings there are lobelike PS
regions (having $n$ and $n+1$ MIs spatially separated given
$n<\bar n < n+1$) at small $\tilde V$, and PL regions outside the
lobes. In the PL case, when $\bar n>1$ and goes across half
integers (dash-dotted line), the ground state suddenly switches to
another subspace of opposite parity and occupations, from a PL
being particle doped from the $n$-filling MI to one being hole
doped from the ($n+1$)-filling MI [denoted as PL($n$,$n+2$) and
PL($n+1$,$n-1$), respectively]. (This continuity in density but
discontinuity in parity could invalidate the local-density
approximation, which is commonly used to profile a trapped system
from the uniform phase diagram\cite{DeMarco05}.) Such parity
boundary meets the tip of the PS lobes at a triple point $\tilde
V^*={A_d}{2^{ - 1/d}}$, which is a function of dimension but
independent of filling. At $\bar n<1$, there is no such triple
point due to the lack of hole-doped PL from the $n=1$ MI. The PL
here is always of even parity [PL(0,2)].

The scaling $\tilde V= V/(UL)$ indicates that the PL will
eventually disappear in the limit of $L \to \infty$, except in the
free case ($U=0$) when it is the PS state instead of the PL one
that disappears. On the other hand, the PL phase is always allowed
in a finite-sized or a trapped system and will exist at the
interphase between different MI plateaus as will be discussed
below. In addition, the rich structure in the $\bar n$--$\tilde V$
diagram implies a nontrivial interplay when all $t$, $U$ and $V$
are non-zero. In Sec.~\ref{sec:numerical}, we numerically study
and present results for a general trapped system using quantum
Monte Carlo and Gutzwiller mean-field methods.

\section{numerical simulations ($t\neq 0$)}\label{sec:numerical}

In this section we apply two commonly used methods to do
calculations for a trapped system (a natural setup in cold-atom
experiments), at various values of the parity coupling in one and
two dimensions. In a harmonic trap, the system has a local
chemical potential of the form
\begin{eqnarray}
{\mu _i} = {\mu _0} - \frac{K}{2}{\bf{r}}_i^2,
\end{eqnarray}
where $\mu_0$ is the global chemical potential, $K$ is the trap's
curvature, and ${\bf{r}}_i$ is the position of site $i$ from the
trap center.

\begin{figure}[t]
\centering
   \includegraphics[width=6.8cm]{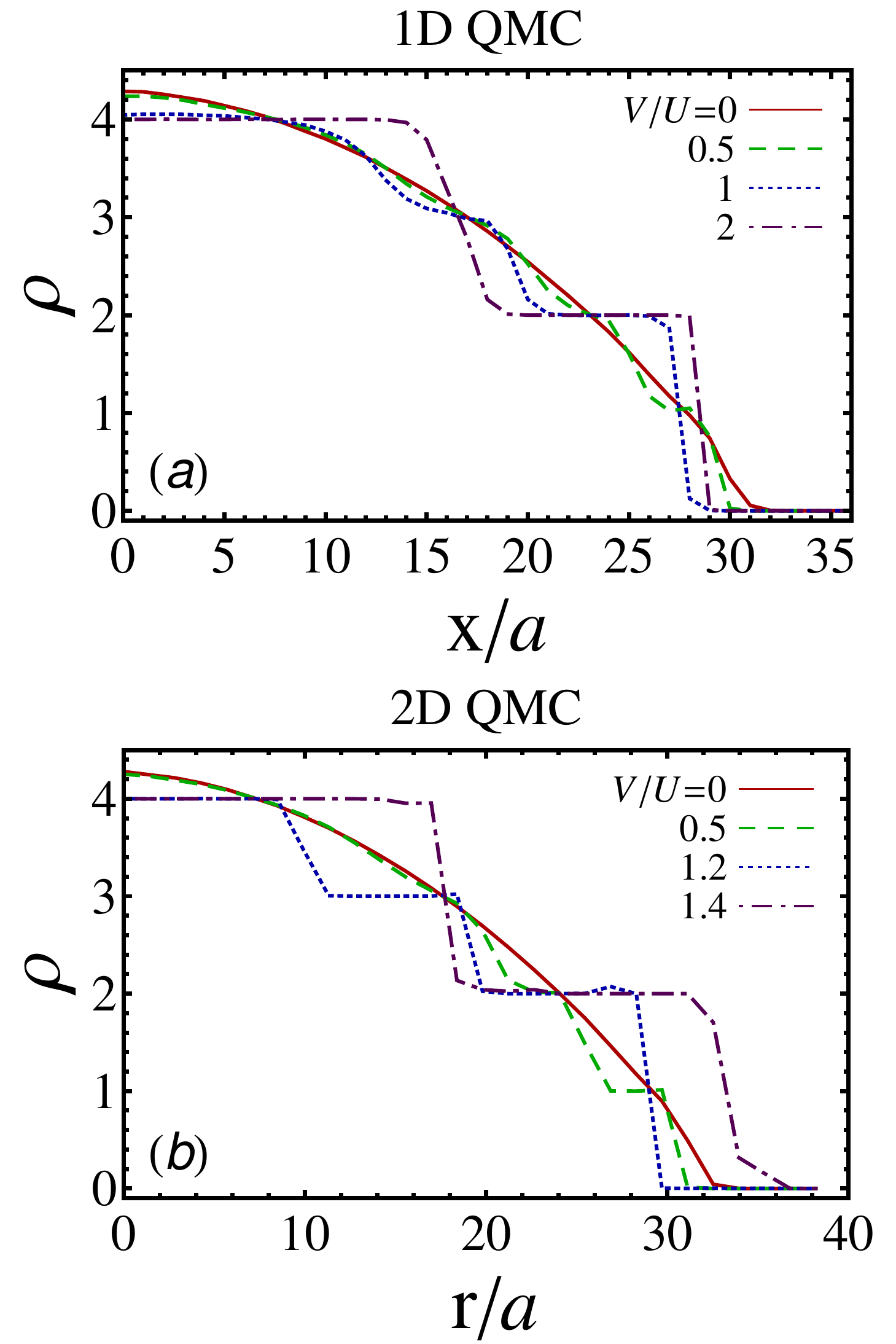}
  \caption{(Color online) Density profiles $\rho$ obtained from the
quantum Monte Carlo (QMC) method at various $V$. (a)
One-dimensional (1D) profiles from the trap center to the edge, at
$V/U=0$, $0.5$, $1$, and $2$. The model parameters are $t/U=0.2$,
$\mu_0/U=3.5$, $K/U=0.0075/a^2$, where $a$ is the lattice spacing,
and the temperature $T/U=0.002$. (b) Two-dimensional (2D) profiles
along the diagonal of a square lattice, from the trap center to
the edge at $V/U=0$, $0.5$, $1.2$, and $1.4$. The model parameters
are $t/U=0.1$, $\mu_0/U=3.5$, $K/U=0.0066/a^2$, and $T/U=0.002$.
In both panels the data are presented in solid, dashed, dotted,
and dot-dashed curves respectively for increasing $V$.}
        \label{fig:f03}
        \vspace{-0.5cm}
\end{figure}

First, we use the quantum Monte Carlo (QMC) method and, in
particular, the stochastic series expansion (SSE)
algorithm\cite{Alet05b,Alet05a,Albuquerque07,Bauer11}, which has
been successfully applied on a variety of BH model
studies\cite{Gygi06,*Kollath07,*Wessel07,*Guertler08,*Schmidt08,*Yamamoto09,*Kalz11,*Arrigoni11}.
We plot the density profile $\rho=\left\langle {\hat n}
\right\rangle$ from the trap center in 1D lattices
[Fig.~\ref{fig:f03}(a)] and that along the diagonal of 2D square
lattices [Fig.~\ref{fig:f03}(b)] at a given $t$ that makes the
system a superfluid when $V=0$. In the original BH case ($V=0$),
the superfluid's density profile monotonically and smoothly
decreases to zero from the center to the edge (solid curve),
matching the shape of the trap. In both the 1D and 2D cases, with
an increase in $V$, the system first develops small MI plateaus at
lower integer fillings $\rho=1,2$ (dashed curves), and then the
$\rho=1$ one disappears while the $\rho=2$ one expands, as well as
higher integer plateaus $\rho=3,4$ develop (dotted curve). Finally
only the even-integer MI plateaus $\rho=2,4$ survive at large $V$
(dot-dashed curve). The first development of the integer plateaus
is because a local domain with integer occupation costs less
parity energy than that with fractional one, while the later
survival of only the even plateaus is attributable to a boundary
behavior, minimizing the interphase energy between plateaus, plus
the fact that the density always decreases to zero (even parity)
at the edge of the system. We notice that the 2D profile exhibits
a rotational symmetry, so at large $V$ it does look like a \lq\lq
wedding cake" but with the adjacent MI plateaus differing by two
units.

\begin{figure}[t]
\centering
   \includegraphics[width=6.8cm]{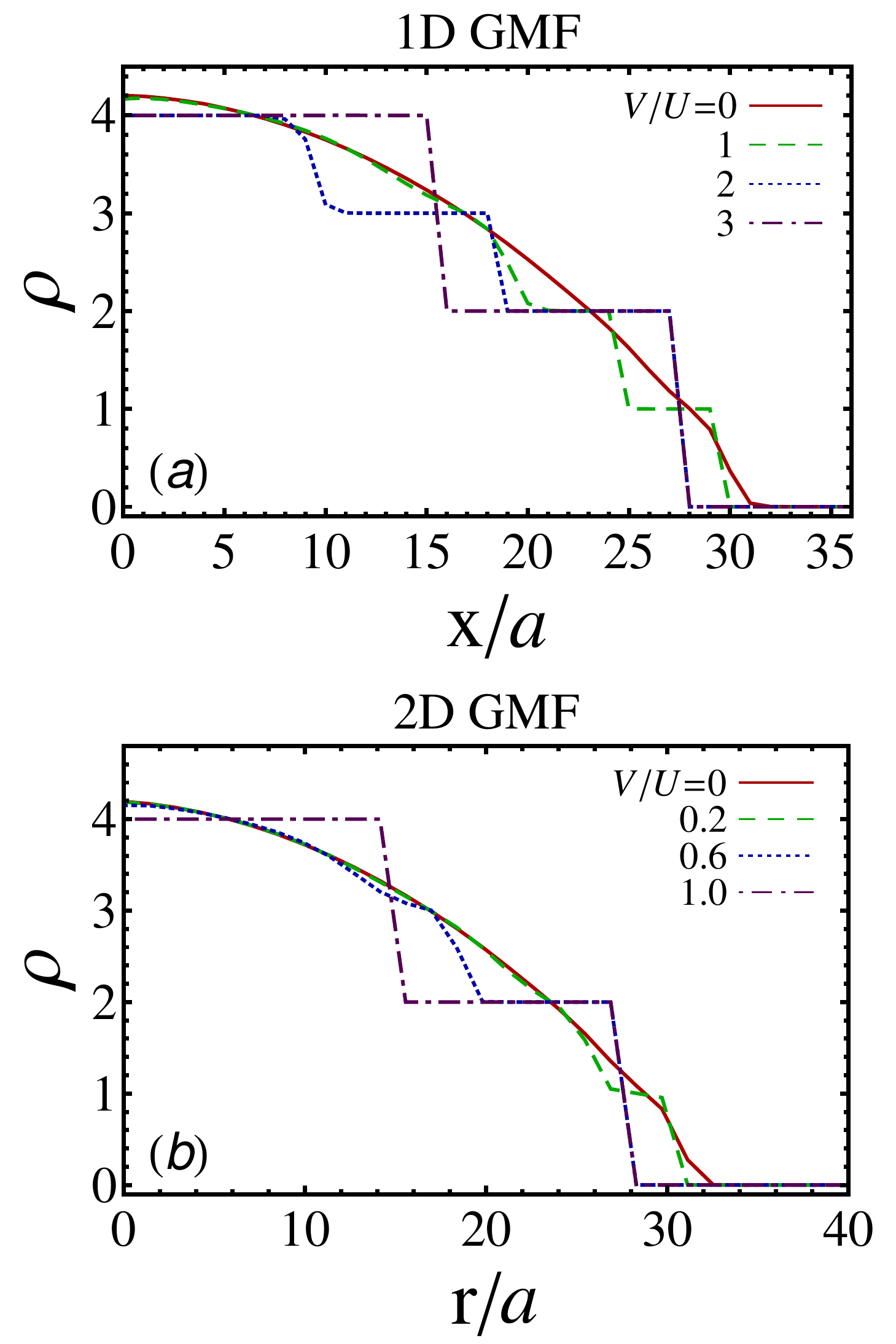}
  \caption{(Color online) Density profiles obtained from the
Gutzwiller mean-field (GMF) method. (a) 1D profiles at $V/U=0$,
$1$, $2$, and $3$. (b) 2D diagonal profiles at $V/U=0$, $0.2$,
$0.6$, and $1$. Data are presented in the same convention as in
Fig.~\ref{fig:f03} and the model parameters are also the same
except for $T=0$.}
        \label{fig:f04}
        \vspace{-0.5cm}
\end{figure}

\begin{figure}[t]
\centering
   \includegraphics[width=6.8cm]{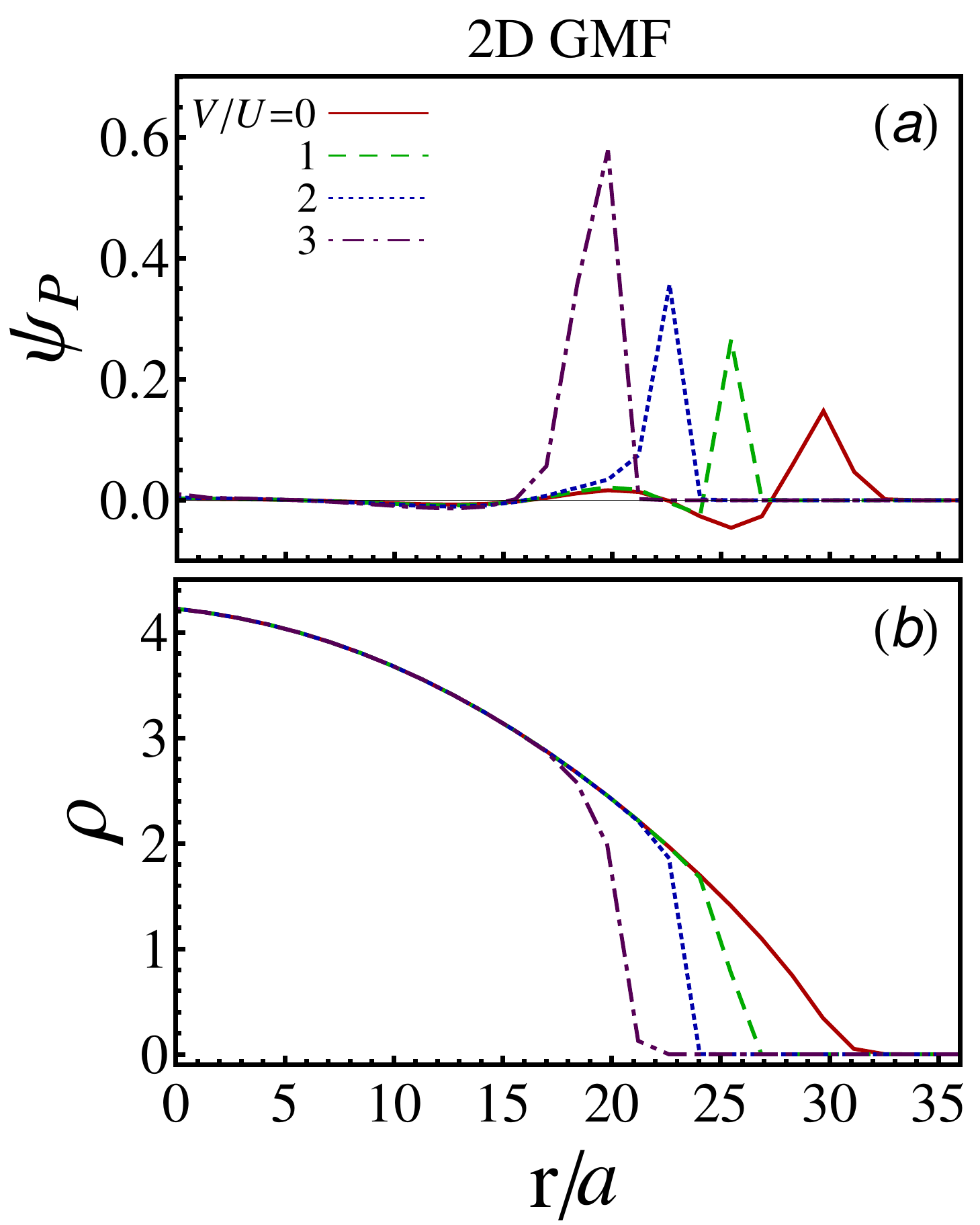}
  \caption{(Color online) (a) Pair-superfluid order ${\psi _P}$ and
(b) the corresponding density profile along the diagonal in 2D
trapped lattices at $V/U=0$, $1$, $2$, and $3$ (solid, dashed,
dotted, and dot-dashed curves respectively). Data are obtained
using GMF with the model parameters being $t/U=0.5$,
$\mu_0/U=2.5$, $K/U=0.0066/a^2$, and $T=0$.}
        \label{fig:f05}
        \vspace{-0.5cm}
\end{figure}

To compare with the QMC results, we use a Gutzwiller mean-field
(GMF) method\cite{Rokhsar91} that has also been widely applied to
a variety of lattice-boson
systems\cite{Albus03,*Pollet04,*Kimura05,*Scarola05,*Miyakawa06,*Buonsante09,*Yamashita09,*Sun09,*Iskin11,*Kimura11,*Saito12,*Pekker12,*Lin12,*Luhmann13,*Yamashita13,*Buonsante13}.
The key assumption of the GMF method is to treat the ground state
of the system as a variational product of all single-site states,
\begin{eqnarray}
\left| {\psi _g^{{\rm{GMF}}}} \right\rangle=\prod\limits_i {\left[
{\sum\limits_{n = 0}^{{n_m}} {{f_{i,n}}\frac{{{{\left( {b_i^\dag }
\right)}^n}}}{{\sqrt {n!} }}} } \right]} \left| {{\rm{vac}}}
\right\rangle,\label{eqn:GMF}
\end{eqnarray}
where $n_m$ is the upper bound of site occupation, $f_{i,n}$ is
the amplitude of $n$ occupation on site $i$, and $ \left|
{{\rm{vac}}} \right\rangle$ denotes the vacuum state. We obtain
the solutions of $\{ f_{i,n} \}$ by minimizing the energy
functional $\left\langle {\psi _g^{{\rm{GMF}}}} \right|H\left|
{\psi _g^{{\rm{GMF}}}} \right\rangle$, subject to the
normalization constraint $\sum\limits_{n = 0}^{{n_m}} {{{\left|
{{f_{i,n}}} \right|}^2}}  = 1$. Here $n_m$ is chosen large enough
such that we observe convergence of the lowest energy state within
the Gutzwiller assumption. Figures \ref{fig:f04}(a) and
\ref{fig:f04}(b) show 1D and 2D density profiles obtained from
GMF, respectively. We see the same four-stage evolution from the
superfluid to the wedding-cake structure of even-occupation MIs,
qualitatively in agreement with the QMC results.

In addition, we investigate a 2D case with larger $t$, in which
the system is always a superfluid ($\langle {\hat b}\rangle \neq
0$ everywhere as long as $\rho \neq 0$) in a range of $V$ of
interest. We find that with the increase in $V$, the outer
superfluid shell develops a strong parity \emph{preference},
$\langle {\hat P} \rangle  \to  \pm 1$, while the inner superfluid
core remains the same state ($|\langle {\hat P} \rangle|  \ll 1$)
as in the $V=0$ case. This parity preference indicates a
pair-doped nature analogous to the PL state in the $t \to 0$ limit
(discussed in Sec.~\ref{sec:phase}), while the concurrence of the
superfluid order suggests the itineracy of particles as pairs,
reminiscent of pair superfluids in multi-species lattice
bosons\cite{Trefzger09,Hu09,Menotti10} or dimer superfluids in BH
systems with attractive
interactions\cite{Diehl10,Bonnes11,Ng11,Chen11}. We identify such
pair-superfluid state by a combined order parameter,
\begin{eqnarray}
{\psi _P} = {| {\langle {\hat b} \rangle } |^2}\langle {\hat P}
\rangle.
\end{eqnarray}
Notice that the sign of $\psi_P$ tells the pair superfluid's
parity being even ($+$) or odd ($-$).

In Fig.~\ref{fig:f05} we plot profiles of $\psi_P$
[\ref{fig:f05}(a)] and $\rho$ [\ref{fig:f05}(b)] obtained from GMF
calculations at a set of increasing $V$ values and $t$ five times
larger than that in Fig.~\ref{fig:f04}(b). At $V=0$, $\psi_P$
shows no parity preference in the central region, due to a smooth
occupation distribution $\left|f_{n} \right |^2$, and slowly
alternates between positive (even) and negative (odd) when the
density becomes dilute near the edge (because the occupation
amplitudes $\{ f_{i,n} \}$ are asymmetrically distributed in $n$,
with a natural cutoff by $n=0$). As $V$ increases, $\psi_P$
develops a strong preference of even parity in the outer shell
region (pair superfluid), leaving the inner core still parity
undefined (regular superfluid). We see that the density of the
outer shell drops much faster when it becomes a pair superfluid,
indicating that the pair superfluid has higher compressibility
($\partial \rho / \partial \mu$) than the regular superfluid. We
point out that the pair superfluid here results from the
co-tunneling of two particles to avoid the flipping of site's
occupation parities. The mechanism is different from the pair
superfluids in other
systems\cite{Trefzger09,Hu09,Menotti10,Diehl10,Bonnes11,Ng11,Chen11},
in which two particles are directly bound by attractive
interactions.

\section{Conclusion}\label{sec:conclusion}
In this paper, we studied an extended Bose-Hubbard model
incorporating a ferromagneticlike coupling between the
nearest-neighbor site-occupation parities. This parity coupling
generates domain-wall energies that compete with the
single-particle tunneling and favors pair doping rather than the
single-particle one that is favored instead by the on-site
repulsion. The interplay between the parity coupling and on-site
repulsion leads to a rich phase diagram of a uniform system at
zero tunneling, characterized by (1) Mott-insulator (MI) phases at
each commensurate filling and (2) phase separation of two MIs when
the on-site interaction dominates or (3) pair-liquid phases when
the parity coupling dominates at incommensurate fillings. In the
finite-tunneling inhomogeneous case, we have obtained both quantum
Monte Carlo and Gutzwiller mean-field results that agreeably show
an evolution of a trapped system from a superfluid to a
wedding-cake-structure of MIs with the same parity (or the
occupation jumping by 2), as the parity coupling increases. In a
relatively large tunneling and large-parity coupling regime, the
trapped system exhibits a structure with the inner bulk being a
single-particle superfluid as usual and the outer shell being an
exotic pair superfluid.

Considering the realization of our model, we have shown that it
effectively describes the behavior of the UMFs in a system of
imbalanced fermionic superfluids in 2D optical lattices of tubes,
which has been realized\cite{Liao10} and is being studied in
ongoing experiments\cite{Hulet13}. The transition between a
superfluid and MIs of the same parity in our model corresponds to
that between compressible and incompressible states of UMFs in the
tubular system with a strong oscillatory pairing
order\cite{Sun12}. Such transition can be detected by the response
of the UMFs to optical-lattice
modulation\cite{Stoferle04,Jordens08} or by the momentum
distribution of UMFs in a time-of-flight
experiment~\cite{Kinoshita04,Swanson12,Lu12,Bolech12}. The
pair-superfluid state in our model corresponds to the compressible
state of itinerant UMF pairs in the tubular system, reminiscent of
a triplet-pair superfluid. In addition, a truncated version of our
model can be mapped to an anisotropic bilinear-biquadratic spin
model, in which the occupation-parity coupling is associated with
biquadratic spin coupling. Such mapping provides a potential
realization for exploring our model in spin systems and would also
stimulate interest in the physics of these two models.

\section*{Acknowledgments}
We are grateful to R.~G.~Hulet, Michael Ma and Matthias Troyer for
useful discussions. The QMC calculations have been performed using
the SSE algorithm provided by the ALPS
project\cite{Alet05a,Albuquerque07,Bauer11}. This work is
supported by DARPA-ARO Award No. W911NF-07-1-0464 and by the
University of Cincinnati. Furthermore, C.J.B.~would like to
acknowledge the hospitality of the Aspen Center for Physics,
supported by the NSF under Grant No. PHYS-1066293.

\appendix*
\section{Deriving the model Hamiltonian from imbalanced fermionic superfluids in optical lattices} \label{sec:DBH}
In this Appendix we derive the Hamiltonian of Eq.~(\ref{eqn:Ham0})
as an effective model describing the behavior of UMFs in a system
of an imbalanced fermionic superfluid in a 2D optical lattice of
1D tubes. The Hamiltonian of the tube system comprises three
parts\cite{Sun12,Sun13},
\begin{eqnarray}
{H_0} &=& \int_z {\sum\limits_{{\bf{r}},\sigma } {\hat \psi
_{\sigma {\bf{r}}}^\dag (z)\left( { - \frac{{{\hbar ^2}\partial
_z^2}}{{2m}} + {U_{\sigma {\bf{r}}}}(z) - {{ \mu }_{\sigma }}(z)}
\right){{\hat \psi }_{\sigma {\bf{r}}}}(z)}
},\nonumber\\\\
{H_1} &=& \int_z {{\Delta _{\bf{r}}}(z)\hat \psi _{ \uparrow
{\bf{r}}}^\dag (z)\hat \psi _{ \downarrow {\bf{r}}}^\dag (z) +
{\rm{H}}{\rm{.c}}{\rm{.}}},\\
{H_2} &=& \int_z {\sum\limits_{\left\langle {{\bf{rr'}}}
\right\rangle ,\sigma } {{\mathcal{T}_{{\bf{rr'}}\sigma }}(z)\hat
\psi _{\sigma {\bf{r}}'}^\dag (z){{\hat \psi }_{\sigma
{\bf{r}}}}(z)} },
\end{eqnarray}
where the $\hat z$ direction is along the tubes' axis and $
{\bf{r}}=(x,y)$ denotes tubes indexed in the lattice plane
perpendicular to $\hat z$. The operator ${{\hat \psi^\dag
_{\sigma=\uparrow/\downarrow,{\bf{r}}}(z)}}$ creates a majority
(minority) spin at position $z$ on tube ${\bf{r}}$. The first part
$H_0$ contains the kinetic energy in the $\hat z$ direction, the
Hartree field $U$ and the spin-dependent chemical potential $\mu$
(we assume homogeneity in the $\hat{\bf{r}}$ direction), the
second part $H_1$ includes the superfluid pairing order $\Delta$,
while the third part $H_2$ models the tunneling between
nearest-neighbor tubes $\left\langle {{\bf{rr'}}} \right\rangle$.
The tunneling field $\mathcal{T}$ describes the effective
single-particle tunneling under the influence of the pairing state
of the surrounding tubes\cite{Sun13}. The three coupling fields
$U$, $\Delta$, and $\mathcal{T}$ are introduced after employing a
Bogoliubove--de Gennes (BdG) treatment\cite{DeGennes66} on the
original Hamiltonian of a quartic form; which includes, besides
the single-particle Hamiltonian, a two-body point-contact
interaction and (possible) pair tunneling. The Hamiltonian $H_0 +
H_1 + H_2$ can be solved in a quasiparticle basis, resulting in a
set of $\{ U,\Delta, \mathcal{T}\}$ that self-consistently
describes the equilibrium state of the system (see detailed
discussions in Refs.~[\onlinecite{Sun12}] and [\onlinecite{Sun13}]
as well as a wide application of this treatment on similar
systems~\cite{Mizushima05,Parish07,Liu07,Baksmaty11,Sun11}).

The self-consistent solutions show that in the presence of spin
imbalance, the system can accommodate an oscillatory pairing order
$\Delta$ along the tube axis (the FFLO
state\cite{FF,LO,Giorgini08,Radzihovsky10}). For a single tube or
uncoupled tubes, each UMF is localized around a node of the
pairing order parameter such that the total numbers of nodes and
UMFs are equal. Given the spatially symmetric confinement along
the tube axis, e.g., a harmonic trap, a tube with odd (even)
imbalance has a pairing node (antinode) at the tube center. In
particular, the localized UMFs can be regarded as occupied Andreev
bound states\cite{Yoshida07,Parish07}, resulting in an imbalance
profile like separated solitons\cite{Brazovskii09,Lutchyn11} if
the spatial distribution of the pairing nodes is relatively sparse
or like a ripple\cite{Liu07} if it is dense. (The same signatures
have also been found in isotropic three-dimensional Hubbard
lattices\cite{Loh10} or coupled Hubbard chains\cite{Kim12}.) Both
configurations arise from the same effect of localization and can
be characterized as the spatial concurrence of imbalance peaks and
pairing nodes.

The localization of UMFs suppresses the degrees of freedom in the
$\hat z$ direction and hence allows us to construct an effective
2D lattice model describing the physics of UMFs with only the
$\hat {\bf{r}}$ degree of freedom. We remark that even if UMFs
localize along the $\hat z$ direction, they can still be mobile in
the$\hat {\bf{r}}$ direction, constituting the physics of
interests as discussed below. In such a 2D model the state in the
$\hat z$ direction is taken as a hidden degree of freedom of a
lattice site. (This idea has also been applied on bosonic gases in
tubular lattices\cite{Wirth11,Li13a,Olschlager13,Hebert13,Li13b},
in which the multiorbital wave functions in the tube direction are
incorporated into on-site degrees of freedom for an effective 2D
model across the tubes.) The hidden degrees of freedom here can be
directly related in one-to-one correspondence to the site
occupation number, and since more than single occupation per site
are possible, this suggests a bosonic description for the
modeling. Formally, one can still keep the anticommutation
relation in the $\hat {\bf{r}}$ direction by using a combined
operator
\begin{eqnarray}
{\psi _{ \uparrow {\bf{r}}}}(z) \to {{\hat \eta }_{\bf{r}}}{{\hat
b}_{\bf{r}}},
\end{eqnarray}
where ${\hat b}$ is a bosonic operator accounting for
occupation-related physics and ${\hat \eta }$ is a Majorana
fermion operator carrying the statistical property of the original
fermions. Based on this decoupling, we build our model step by
step below.

First, for coupled tubes, the pairing nodes tend to line up at the
same $z$ positions on each tube\cite{Sun12,Sun13}. The leading
kinetics across the tubes is the transverse tunneling of an UMF
from a pairing node in a tube to the corresponding one in the
nearest neighbor. The one-particle tunneling of minority fermions
is suppressed provided they are all paired (at low temperature).
Because a node is surrounded by other nodes where the pairing is
zero, the tunneling field reduces to a free particle tunneling
strength\cite{Sun13}, $\mathcal{T} \to -t$. Therefore, considering
the order-parameter configuration in $H_1$ that limits the
kinetics in $H_2$ to be only for the UMFs, we obtain the tunneling
term in the model as
\begin{eqnarray}
- t\sum\limits_{\left\langle {{\bf{rr'}}} \right\rangle } {\psi _{
\uparrow {\bf{r'}}}^\dag {\psi _{ \uparrow {\bf{r}}}}} \to  -
t\sum\limits_{\left\langle {{\bf{rr'}}} \right\rangle } {\hat
b_{{\bf{r'}}}^\dag {{\hat \eta }_{{\bf{r'}}}}{{\hat \eta
}_{\bf{r}}}{{\hat b}_{\bf{r}}}}\label{eqn:Ham2a}
\end{eqnarray}

Second, if the coupling between the tubes is weak enough, the
on-site energy is mainly contributed by the on-tube kinetics,
Hartree energy, pairing energy, and chemical potentials, all of
which are determined once the on-site filling (of UMFs) is given.
The self-consistent calculation in Ref.~[\onlinecite{Sun12}] shows
that the on-site energy is nonlinear and can be fitted by a
repulsive interaction $U$ plus a chemical potential $\mu$. As a
result, the on-tube energies $H_0$ and $H_1$ effectively become
the on-site terms of our model as
\begin{eqnarray}
\sum\limits_{\bf{r}} {\frac{U}{2}{{\hat n}_{\bf{r}}}({{\hat
n}_{\bf{r}}} - 1) - \mu {{\hat n}_{\bf{r}}}},\label{eqn:Ham2b}
\end{eqnarray}
where ${{\hat n}_{\bf{r}}} = \hat b_{\bf{r}}^\dag {{\hat \eta
}_{\bf{r}}}{{\hat \eta }_{\bf{r}}}{{\hat b}_{\bf{r}}} = \hat
b_{\bf{r}}^\dag {{\hat b}_{\bf{r}}}$.

Third, the BdG calculations show that the system energy rises if
two nearest-neighbor tubes have different spatial parities of the
oscillatory pairing order parameter. Such energy increase comes
from a drastic mismatch in the order parameters and can be
regarded as a cost of domain-wall formation, exhibiting similar
physics to that in superconducting $\pi$ junctions\cite{Buzdin05}
or weakly imbalanced cold atomic gases\cite{Yoshida07}. Because
the spatial parity of the order parameter is equal to the
occupation parity of the UMFs, one can describe the domain-wall
physics with the occupation-parity coupling between
nearest-neighbor sites in the BH model,
\begin{eqnarray}
-\frac{V}{2}\sum\limits_{\left\langle {{\bf{rr'}}} \right\rangle }
{\left( {{{\hat P}_{{\bf{r'}}}}{{\hat P}_{\bf{r}}}-1}
\right)},\label{eqn:Ham2c}
\end{eqnarray}
where ${{\hat P}_{\bf{r}}} = {( - 1)^{{{\hat n}_{\bf{r}}}}}$ is
the onsite parity operator and $V$ is the unit-length energy of
domain walls.

Finally, by collecting all terms in
Eqs.~(\ref{eqn:Ham2a})--(\ref{eqn:Ham2c}) we obtain an effective
Hamiltonian,
\begin{eqnarray}
&&\sum\limits_{\left\langle {{\bf{rr'}}} \right\rangle } {\left[ {
- t\left( {\hat b_{{\bf{r'}}}^\dag {{\hat \eta
}_{{\bf{r'}}}}{{\hat \eta }_{\bf{r}}}{{\hat b}_{\bf{r}}} +
{\rm{H}}{\rm{.c}}{\rm{.}}} \right) - \frac{V}{2}\left( { {{\hat
P}_{{\bf{r'}}}}{{\hat
P}_{\bf{r}}}-1} \right)} \right]} \nonumber\\
&&+ \sum\limits_{\bf{r}} {\left[ {\frac{U}{2}{{\hat
n}_{\bf{r}}}({{\hat n}_{\bf{r}}} - 1) - \mu {{\hat n}_{\bf{r}}}}
\right]}.\label{eqn:Ham2}
\end{eqnarray}
Because the physical quantities of interest to be addressed here
can be expressed as local static (or equal-time) correlations of
the ground state, not involving the exchange of particles, the
statistics (Majorana sector) factors out. Therefore, by replacing
$\hat b_{{\bf{r'}}}^\dag {{\hat \eta }_{{\bf{r'}}}}{{\hat \eta
}_{\bf{r}}}{{\hat b}_{\bf{r}}}$ with $\hat b_{{\bf{r'}}}^\dag
{{\hat b}_{\bf{r}}}$ (and allowing the chemical potential to vary
in space), we arrive at the Hamiltonian of Eq.~(\ref{eqn:Ham0}).
Considering a typical experimental setup\cite{Liao10}, we obtain
$U \sim 0.1\epsilon_b$, $t/U\sim 0.1$--$10\%$ in the range of
optical lattice depth being 12--3$E_R$, and $V/U\sim 10$ (regime
of the wedding-cake structure of even-filling MIs), where
$\epsilon_b$ is the binding energy of a pair of fermions and $E_R$
is the recoil energy. One can also expect a high tunability of $U$
and $V$ via the tuning of the Feshbach-resonant\cite{Chin10}
interactions and the overall trapping potential used in the
experiments.

\end{document}